# Influence of fractional composition of crystallite grains on the dark conductivity in fully crystallized undoped microcrystalline silicon

Sanjay K. Ram*,a, Satyendra Kumar*,b and P. Roca i Cabarrocas!

*Department of Physics, Indian Institute of Technology Kanpur, Kanpur-208016, India
!LPICM, UMR 7647 - CNRS - Ecole Polytechnique, 91128 Palaiseau Cedex, France

Improvement in film growth technology requires a knowledge of the correlation between microstructural and deposition parameters with electrical properties in hydrogenated microcrystalline Si ($\mu$c-Si:H) films. Our study indicates that fractional compositions of the constituent crystallite grains in fully crystallized undoped $\mu$c-Si:H films is a unique microstructural feature that defines the film microstructure and can be well correlated to the electrical transport properties as well.

Plasma deposited hydrogenated microcrystalline silicon ($\mu$c-Si:H) has become the subject of extensive study as it offers the possibilities of high carrier mobility[1,2] and stability against Staebler-Wronski effects.[3,4] The important microstructural feature of plasma deposited $\mu$c-Si:H films is the presence of crystallite grains that conglomerate to form larger columnar features growing perpendicular to the substrate.[5,6,7] Disordered or amorphous phase and voids populate the boundaries of the grains and columns.[8,9] Electronic transport in $\mu$c-Si:H films has been variously claimed to be analogous to that observed in $a$-Si:H films[10] and poly-Si films. Recent advances in film growth technologies have led to reproducible highly/ fully crystallized $\mu$c-Si:H films having no distinguishable amorphous phase,[11,12,13] in which the time-honored concepts regarding the roles of changing crystallinity with film growth and amorphous phase in the system become rather inconsequential. In such a situation, it is not clear which easily identifiable and measurable microstructural attribute influencing the electrical properties can be useful for parameterizing these properties as well. Recently, the investigation into the electrical transport mechanisms and routes has revealed the importance of percolation conduction through a network of connected disordered tissue encapsulating the columns in fully-crystalline undoped $\mu$c-Si:H.[14] The role of crystallite grain boundaries (GB) in potential barriers and film transport properties has been explored in previous works.[15,16,17]

Many studies have been devoted to understand how variation in deposition parameters can influence transport properties and can be used to tune them. However, the deposition parameters themselves do not have a unique causal relationship with any specific film microstructure, as the same microstructure can be achieved by the adjustment of one or more deposition parameters. Rather, there must exist one or more unifying microstructural characteristics resulting from the alteration in deposition parameters, which have a physical influence on the electrical properties. Earlier works have reported on the correlation between electronic transport properties and the microstructural parameters of $\mu$c-Si:H films such as crystallinity,[18,19] film thickness,[19,20,21] crystallite size,[22,23] crystalline orientation[24,25,26] and conglomerate crystallite sizes.[27] These transport properties have been explained using models that invoke potential barriers at grain boundaries[15,28] and percolation.[16,17,22,29] Again, each of these parameters has only a secondary influence on the electrical properties through its primary effect on the film microstructure. In this present work, we have explored the relationship between fractional compositions of the constituent crystallite grains and the electrical transport properties in fully crystallized undoped single phase $\mu$c-Si:H films. Our study reveals that the fractional composition of constituent crystallite grains is reflective of the film microstructure and morphology; and though not directly implicated in relation to transport routes, is an easily measurable parameter (using ellipsometry or Raman spectroscopy) that is useful to predict electrical transport behavior.

In this work, we carried out a detailed study of the film microstructure correlative with different deposition parameters, in context of the consequent electrical properties. For this we have prepared undoped $\mu$c-Si:H films by depositing at low substrate temperature ($T_s \leq 200$°C) in a parallel-plate glow discharge plasma enhanced chemical vapor deposition system operating at a standard rf frequency of 13.56 MHz, using high purity $SiF_4$, Ar and $H_2$ as feed gases. We systematized our work by studying the influence of varying any one of the two major deposition parameters, namely, gas flow ratios ($R=SiF_4/H_2$; $SiF_4$=1sccm and $H_2$ dilution range 1-20sccm) and $T_s$ (100-250°C) on the film microstructure, for different film thickness (~50-1200nm), thereby creating a series of samples. We employed Raman scattering (RS), spectroscopic ellipsometry (SE), X-ray diffraction (XRD), and atomic force microscopy (AFM) for structural investigations. Many of the $\mu$c-Si:H films used in this study have been characterized by the time resolved microwave conductivity (TRMC) measurements as well. High crystallinity of all the samples was confirmed by RS and SE measurements. SE data shows a crystalline volume fraction >90% from the initial stages of growth, with the rest being density deficit having no amorphous phase, and a reduced incubation layer thickness. The fractional composition of the films educed from SE data shows crystallite grains of two distinct sizes, which is corroborated by the deconvolution of RS profiles using a bimodal size distribution[30,31] of large crystallite grains (LG ~70-80 nm) and small crystallite grains (SG ~6-7 nm). The XRD results have demonstrated the LG and SG to be having different orientations. The presence of a size distribution in the surface conglomerate grains was also established by AFM. However, there is a significant variation in the percentage volume fraction of the constituent LG and SG with film growth. Preferential orientation in (400) and (220) directions was achieved by optimizing the deposition conditions leading to smooth top surfaces (surface roughness < 3 nm).[32,33]

[a]skram@iitk.ac.in (Corresponding author: S.K. Ram).
[b]satyen@iitk.ac.in

Coplanar dark conductivity $\sigma_d(T)$ measurements were carried out from ~300K to ~450K on these well-characterized annealed samples having a variety of microstructures, and studied in context of deposition parameters such as deposition time, $H_2$ dilution, and substrate temperature, as they are known to influence the film microstructure. At above room temperature, $\sigma_d(T)$ of all the $\mu$c-Si:H films having different microstructures, prepared under different deposition conditions, follows Arrhenius type thermally activated behavior: $\sigma_d = \sigma_0 e^{-E_a/kT}$, where $\sigma_o$ is known as conductivity pre-factor and $E_a$ as activation energy. In all the samples, with an increase in film thickness, $\sigma_d$ increases and $E_a$ decreases, irrespective of the deposition parameter values.[34] Our structural studies evince a profound effect of $H_2$ dilution on film crystallinity and crystallite orientation.[33] However, the effect of $H_2$ dilution on electrical transport properties is not systematic, and rather is related to the microstructural and morphological changes that it brings about in the film.[34] The influence of substrate temperature on film microstructure results in a lowering of $E_a$ for the films deposited at higher substrate temperatures.

Figure 1(a) shows a summary of $\sigma_d$ and $E_a$ data plotted as a function of film thickness (films belonging to all the three thickness series of different $R=1/1$, 1/5 and 1/10). For a systematic understanding, the results can be grouped into three different thickness zones as marked in this figure. In the first zone describing lower film thickness, the $\sigma_d$ is low in magnitude and increases with an increase in the film thickness. However, the value of film thickness after which the increase of $\sigma_d$ slows down is ~250 nm for the samples prepared under $R=1/10$, while it is about 400 nm for the samples prepared at $R=1/1$. On the other hand, $E_a$ remains almost constant at ~ 0.55 eV. Second zone in the intermediate thickness range shows increase in $\sigma_d$ by several orders of magnitude accompanied by a decrease in $E_a$. Third zone shows a near saturation in $\sigma_d$ and $E_a$.

The percentage volume fraction of constituent large grains ($F_{CL}$%) and small grains ($F_{CF}$%) in bulk of the films (as described above) is plotted in Fig.1(b) on the same scale. The $F_{CL}$% increases with film growth with a complementary decrease in $F_{CF}$%. The evolving large crystallite grains and columns associated with the film growth influence the electrical transport behavior. We will now consider the variation of $F_{CL}$% and surface roughness with film growth (Fig. 1(c)) together to compositely view the morphological changes that occur with film growth resulting in different electrical transport behaviors. We note that the variation of $F_{CL}$% with thickness can also be grouped into the same three zones, as for the electrical transport parameters (Fig. 1b).

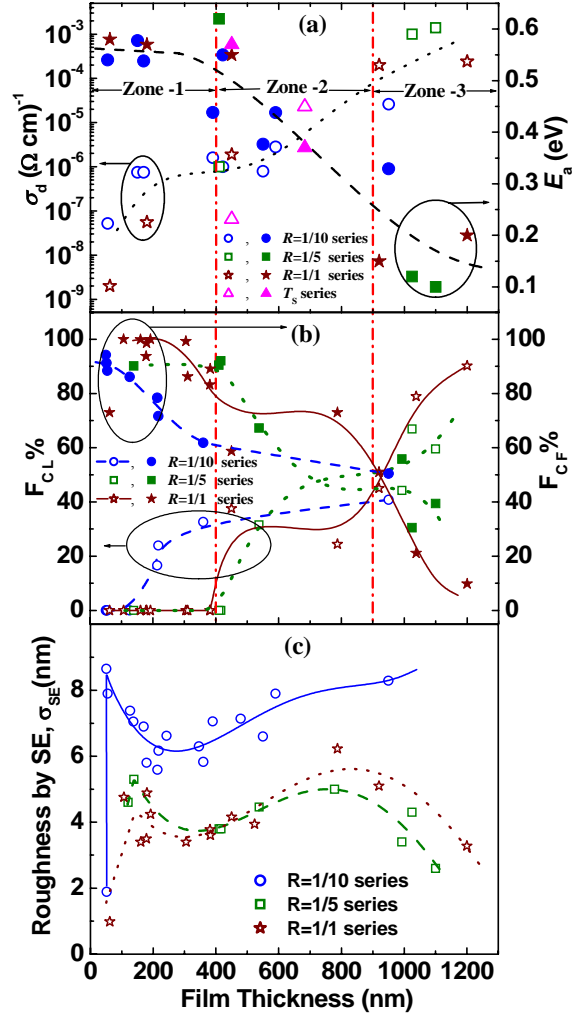

FIG. 1. (a) Variation of $\sigma_d$ (open symbols) and $E_a$ (filled symbols) with film thickness. Samples are of different thickness series, prepared with $R=SiF_4/H_2 =1/1,1/5$ and 1/10 at constant $T_s=200$ °C and at varying $T_s$ at $R=1/5$); (b) Variation of percentage volume fraction of large grains ($F_{CL}$%) (open symbols) and small grains ($F_{CF}$%) (filled symbols) with film thickness. Samples are the same series as in part (a); (c) Dependence of top surface roughness layer ($\sigma_{SE}$, calculated by in-situ SE) on film thickness of the same samples. Here lines are shown as guide to the eye.

In zone-1, the $F_{CL}$% increases from 0–35% due to which an increase is seen in $\sigma_d$. Here, the samples mainly contain SG ~ ≤10 nm in size. Some of the thinner samples may also contain amorphous silicon tissue in the incubation layer. Having a large number density of SG with a distribution of crystallites sizes leads to higher structural disorder even if there is no appreciable presence of an amorphous silicon tissue. Considering the surface roughness in this zone, initially the roughness is found to increase (film thickness<50-70 nm), followed by a decrease until thickness ~300 nm. The rise in roughness is because of small grain formation or crystallization occurring in the initial growth phase, while the decrease heralds the start of coalescences of these small grains to form bigger crystallites. The $\sigma_d$ in this zone is low, and the $E_a$ around room temperature is high (~0.55 eV). In our opinion, this type of $\mu$c-Si:H material has been treated as homogeneous and analogous to $a$-Si:H in literature. In zone-2, $F_{CL}$% is almost constant in the range of ~30- 40%. The samples lying in this zone were found to have large variation in film morphology and detailed microstructure as indicated by AFM and RS analysis. Here we observe a second stage of rise in roughness (from 300-750 nm) denoting crystallite column formation while the fractional compositions of SG and LG remain almost constant, and the columns start coalescing with each other during the coalescence phase. The electrical transport in this zone is strongly influenced by the formation of columnar boundaries and hence large variations in $\sigma_d$ and its $E_a$ are seen. Zone-3 represents films constituted primarily of tightly packed crystallites having LG ($F_{CL}$% >50%) and SG with no amorphous silicon tissue. The AFM studies of these samples show that the conglomerate columns have an average width ~300-400nm. Preferred orientation is present in these films. The void fraction (density deficit) in these samples is negligible. Once the coalescence phase of zone-2 is over, the film growth again accelerates, and simultaneously roughness starts decreasing. Thus, the roughness in this zone approaches a constant value when the film growth reaches a steady state phase.

We note here that the variation of $\sigma_d$ and $E_a$ with respect to thickness or $R$ or $T_s$ can be well understood only when viewed in the perspective of $F_{CL}$%. Thus, it appears that thickness is not a rational parameter in itself, rather, it is the microstructure specific to a thickness zone that is an important determinant of the zone's characteristic properties, and not the thickness per se.

In discussing transport mechanisms in $\mu$c-Si:H we need to distinguish between highly doped samples and undoped or unintentionally doped samples. In heavily doped $\mu$c-Si:H material current route follows through crystallites/ columns and transport properties can be understood by well established GBT models, whereas recent experimental evidence strongly suggests a dominant role of the disordered Si tissue of the boundaries encapsulating the crystallite columns in electrical transport in fully crystalline single phase undoped $\mu$c-Si:H material,[14] and thus endorsing a band tail transport. Although the trend of the electrical transport behavior is evident from Fig 1(a and b), it would still be desirable to see how the material parameters like conductivity prefactor ($\sigma_0$) and $E_a$ correlate with $F_{CL}$%. $\sigma_0$ is an indicator of the amount of statistical shift in $E_f$ and mobility edges. We have plotted $\sigma_0$ and $E_a$ data as a function of $F_{CL}$% in Fig. 2(a). In the range where $F_{CL}$% <30%, $\sigma_0$ and $E_a$ are constant. Here the material consists mainly of SG with an increased number of SG boundaries. Therefore, the question of formation of potential barrier (i.e. transport through crystallites) does not arise because the large number of defect/trap sites compared to free electrons and small size of crystallites will result in a depletion width that is sufficiently large to become greater than the grain size, causing the entire grain to be depleted. Therefore, the transport will be governed by the band tail transport. In the range where $F_{CL}$% varies from 30% to 45%, there is a sharp drop in $\sigma_0$ and $E_a$. This is when many morphological changes are occurring during the film growth, and there is a change in the transport routes. The improvement in film microstructure leads to a delocalization of the tail states causing the $E_f$ to move towards the band edges, closer to the current path at $E_c$. The statistical shift of $E_f$ depends on the temperature and the initial position of $E_f$, and when the $E_f$ is closer to any of the tail states and the tail states are steep, its statistical shift is rapid and marked.

In the range where $F_{CL}$% >45%, $\sigma_0$ shows a rising trend and the fall in $E_a$ is slowed down. Cursorily, it appears that the apparent low values of $E_a$ is GB barrier height formed at the interface between neighboring crystallites/ columns, and the appearance of reduction in $E_a$ is a reduction in barrier height with film growth/ increasing $F_{CL}$%, in a manner similar to when the mobility-barrier height variation is seen to match the conductivity-$E_a$ with increase in doping. However, the calculated values of free electron concentrations (from $\mu_{TRMC}$, $\sigma_d$ and $E_a$ data) do not suggest the possibility of unintentional doping achieving such a high value of background doping concentration. In this zone, a higher $F_{CL}$% and large size of columns result in less columnar boundaries, a well-established network of such interconnected boundaries, and thus higher conductivity (rise in $\sigma_0$). Considering transport through the encapsulating disordered tissue, a band tail transport is mandatory. The large columnar microstructure results in a long range ordering which is sufficient to delocalize an appreciable range of states in the tail state distribution. In addition, higher density of available free carriers and low value of defect density can cause a large increase in negatively charged dangling bond state density together with a decrease in positively charged dangling bond states in the gap, which results in a lower DOS near the conduction band edge and can create a possibility of a steeper conduction band tail.[35,36,37] The mobility values obtained from TRMC ($\mu_{TRMC}$) on some of the samples are shown as a function of $F_{CL}$% in Fig. 2(b). TRMC is known to measure the mobility of carriers within the grains.[38] The graph shows that the mobility of carriers within crystallites increases by over an order of magnitude as the percentage of $F_{CL}$% increases, indicating a low defect density within large crystallite grains.

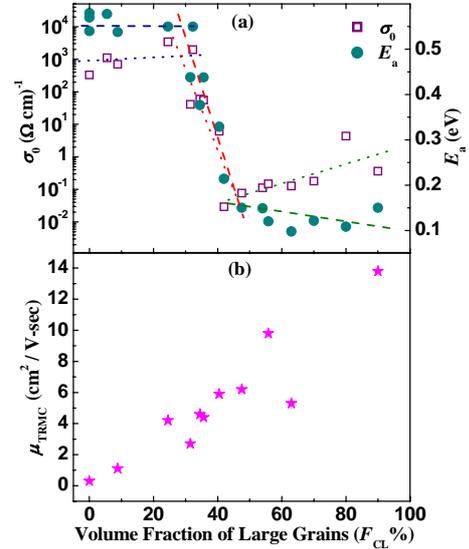

FIG. 2. (a) Variation of $\sigma_0$ (open squares) and $E_a$ (filled circles) with $F_{CL}$%. Here left Y-axis depicts $\sigma_0$ and right Y-axis depicts $E_a$. (b) Dependence of mobility values determined from TRMC measurement ($\mu_{TRMC}$) on $F_{CL}$%.

The role of varying crystallinity in electrical transport behavior and transport routes has been well explored in literature. Our present study offers to correlate a microstructural attribute, namely, the fractional composition of constituent crystalline grains, with the electrical transport behavior in fully crystallized undoped $\mu$c-Si:H films, where there is no appreciable change in overall crystallinity with film growth. Our study indicates that microstructural changes underlie the variation of electrical transport behavior observed with change in thickness, and the $F_{CL}$% can be useful in parameterizing the film microstructure and the thermally activated carrier transport linked thereto. This correlation is applicable in all the three zones of film microstructure having distinct microstructural and morphological characteristics and associated electrical properties, depicting a wide range of $\mu$c-Si:H materials. These results comprehensively string together the scattered and varying features of film microstructure and electrical transport properties of the $\mu$c-Si:H system; and the usefulness of $F_{CL}$% as such a parameter as we have proposed, can be extended to any fully crystalline heterogeneous crystalline composite material where a size distribution of crystallite grains exist.